\def\spacingset#1{\renewcommand{\baselinestretch}{#1}\small\normalsize}
\documentstyle[12pt]{article}
\begin{document}
\newcommand{\nc}{\newcommand}
\nc{\rnc}{\renewcommand}
\nc{\nt}{\newtheorem}
\nc{\be}{\begin}
\nc{\erf}[1]{$\ (\ref{#1}) $}
\nc{\rf}[1]{$\ \ref{#1} $}
\nc{\lb}[1]{\mbox {$\label{#1}$} }
\nc{\hr}{\hrulefill}

\nc{\eq}{\begin{equation}}
\nc{\en}{\end{equation}}
\nc{\eqa}{\begin{eqnarray}}
\nc{\ena}{\end{eqnarray}}

\nc{\ra}{\rightarrow}
\nc{\la}{\leftarrow}
\nc{\da}{\downarrow}
\nc{\ua}{\uparrow}
\nc{\Ra}{\Rightarrow}
\nc{\La}{\Leftarrow}
\nc{\Da}{\Downarrow}
\nc{\Ua}{\Uparrow}

\nc{\uda}{\updownarrow}
\nc{\Uda}{\Updownarrow}
\nc{\lra}{\longrightarrow}
\nc{\lla}{\longleftarrow}
\nc{\llra}{\longleftrightarrow}
\nc{\Lra}{\Longrightarrow}
\nc{\Lla}{\Longleftarrow}
\nc{\Llra}{\Longleftrightarrow}

\nc{\mt}{\mapsto}
\nc{\lmt}{\longmapsto}
\nc{\lt}{\leadsto}
\nc{\hla}{\hookleftarrow}
\nc{\hra}{\hookrightarrow}
\nc{\lgl}{\langle}
\nc{\rgl}{\rangle}

\nc{\stla}{\stackrel{d}{\la}}
\nc{\pard}{\partial \da}
\nc{\gdot}{\circle*{0.5}}


\rnc{\baselinestretch}{1.2}      
\nc{\bl}{\vspace{1ex}}           
\rnc{\theequation}{\arabic{section}.\arabic{equation}}  
\newcounter{xs}
\newcounter{ys}
\newcounter{os}
\nt{thm}{Theorem}[section]
\nt{dfn}[thm]{Definition}
\nt{pro}[thm]{Proposition}
\nt{cor}[thm]{Corollary}
\nt{con}[thm]{Conjecture}
\nt{lem}[thm]{Lemma}
\nt{rem}[thm]{Remark}
\nc{\Poincare}{\mbox {Poincar$\acute{\rm e}$} }

\nc{\bA}{\mbox{${\bf A}$\ }}
\nc{\bB}{\mbox{${\bf B}$\ }}
\nc{\bC}{\mbox{${\bf C}$\ }}
\nc{\bD}{\mbox{${\bf D}$\ }}
\nc{\bE}{\mbox{${\bf E}$\ }}
\nc{\bF}{\mbox{${\bf F}$\ }}
\nc{\bG}{\mbox{${\bf G}$\ }}
\nc{\bH}{\mbox{${\bf H}$\ }}
\nc{\bI}{\mbox{${\bf I}$\ }}
\nc{\bJ}{\mbox{${\bf J}$\ }}
\nc{\bK}{\mbox{${\bf K}$\ }}
\nc{\bL}{\mbox{${\bf L}$\ }}
\nc{\bM}{\mbox{${\bf M}$\ }}
\nc{\bN}{\mbox{${\bf N}$\ }}
\nc{\bO}{\mbox{${\bf O}$\ }}
\nc{\bP}{\mbox{${\bf P}$\ }}
\nc{\bQ}{\mbox{${\bf Q}$\ }}
\nc{\bR}{\mbox{${\bf R}$\ }}
\nc{\bS}{\mbox{${\bf S}$\ }}
\nc{\bT}{\mbox{${\bf T}$\ }}
\nc{\bU}{\mbox{${\bf U}$\ }}
\nc{\bV}{\mbox{${\bf V}$\ }}
\nc{\bW}{\mbox{${\bf W}$\ }}
\nc{\bX}{\mbox{${\bf X}$\ }}
\nc{\bY}{\mbox{${\bf Y}$\ }}
\nc{\bZ}{\mbox{${\bf Z}$\ }}
\nc{\cA}{\mbox{${\cal A}$\ }}
\nc{\cB}{\mbox{${\cal B}$\ }}
\nc{\cC}{\mbox{${\cal C}$\ }}
\nc{\cD}{\mbox{${\cal D}$\ }}
\nc{\cE}{\mbox{${\cal E}$\ }}
\nc{\cF}{\mbox{${\cal F}$\ }}
\nc{\cG}{\mbox{${\cal G}$\ }}
\nc{\cH}{\mbox{${\cal H}$\ }}
\nc{\cI}{\mbox{${\cal I}$\ }}
\nc{\cJ}{\mbox{${\cal J}$\ }}
\nc{\cK}{\mbox{${\cal K}$\ }}
\nc{\cL}{\mbox{${\cal L}$\ }}
\nc{\cM}{\mbox{${\cal M}$\ }}
\nc{\cN}{\mbox{${\cal N}$\ }}
\nc{\cO}{\mbox{${\cal O}$\ }}
\nc{\cP}{\mbox{${\cal P}$\ }}
\nc{\cQ}{\mbox{${\cal Q}$\ }}
\nc{\cR}{\mbox{${\cal R}$\ }}
\nc{\cS}{\mbox{${\cal S}$\ }}
\nc{\cT}{\mbox{${\cal T}$\ }}
\nc{\cU}{\mbox{${\cal U}$\ }}
\nc{\cV}{\mbox{${\cal V}$\ }}
\nc{\cW}{\mbox{${\cal W}$\ }}
\nc{\cX}{\mbox{${\cal X}$\ }}
\nc{\cY}{\mbox{${\cal Y}$\ }}
\nc{\cZ}{\mbox{${\cal Z}$\ }}

\nc{\rightcross}{\searrow \hspace{-1 em} \nearrow}
\nc{\leftcross}{\swarrow \hspace{-1 em} \nwarrow}
\nc{\upcross}{\nearrow \hspace{-1 em} \nwarrow}
\nc{\downcross}{\searrow \hspace{-1 em} \swarrow}
\nc{\prop}{| \hspace{-.5 em} \times}
\nc{\wh}{\widehat}
\nc{\wt}{\widetilde}
\nc{\nonum}{\nonumber}
\nc{\nnb}{\nonumber}
 \nc{\half}{\mbox{$\frac{1}{2}$}}
\nc{\Cast}{\mbox{$C_{\frac{\infty}{2}+\ast}$}}
\nc{\Casth}{\mbox{$C_{\frac{\infty}{2}+\ast+\frac{1}{2}}$}}
\nc{\Casm}{\mbox{$C_{\frac{\infty}{2}-\ast}$}}
\nc{\Cr}{\mbox{$C_{\frac{\infty}{2}+r}$}}
\nc{\CN}{\mbox{$C_{\frac{\infty}{2}+N}$}}
\nc{\Cn}{\mbox{$C_{\frac{\infty}{2}+n}$}}
\nc{\Cmn}{\mbox{$C_{\frac{\infty}{2}-n}$}}
\nc{\Ci}{\mbox{$C_{\frac{\infty}{2}}$}}
\nc{\Hast}{\mbox{$H_{\frac{\infty}{2}+\ast}$}}
\nc{\Hasth}{\mbox{$H_{\frac{\infty}{2}+\ast+\frac{1}{2}}$}}
\nc{\Hasm}{\mbox{$H_{\frac{\infty}{2}-\ast}$}}
\nc{\Hr}{\mbox{$H_{\frac{\infty}{2}+r}$}}
\nc{\Hn}{\mbox{$H_{\frac{\infty}{2}+n}$}}
\nc{\Hmn}{\mbox{$H_{\frac{\infty}{2}-n}$}}
\nc{\HN}{\mbox{$H_{\frac{\infty}{2}+N}$}}
\nc{\HmN}{\mbox{$H_{\frac{\infty}{2}-N}$}}
\nc{\Hi}{\mbox{$H_{\frac{\infty}{2}}$}}
\nc{\Ogast}{\mbox{$\Omega_{\frac{\infty}{2}+\ast}$}}
\nc{\Ogi}{\mbox{$\Omega_{\frac{\infty}{2}}$}}
\nc{\Wedast}{\bigwedge_{\frac{\infty}{2}+\ast}}

\nc{\Fen}{\mbox{$F_{\xi,\eta}$}}
\nc{\Femn}{\mbox{$F_{\xi,-\eta}$}}
\nc{\Fp}{\mbox{$F_{0,p}$}}
\nc{\Fenp}{\mbox{$F_{\xi',\eta'}$}}
\nc{\Fuv}{\mbox{$F_{\mu,\nu}$}}
\nc{\Fuvp}{\mbox{$F_{\mu',\nu'}$}}
\nc{\cpq}{\mbox{$c_{p,q}$}}
\nc{\Drs}{\mbox{$\Delta_{r,s}$}}
\nc{\spq}{\mbox{$\sqrt{2pq}$}}
\nc{\Mcd}{\mbox{$M(c,\Delta)$}}
\nc{\Lcd}{\mbox{$L(c,\Delta)$}}
\nc{\Wxv}{\mbox{$W_{\chi,\nu}$}}
\nc{\vxv}{\mbox{$v_{\chi,\nu}$}}
\nc{\dd}{\mbox{$\widetilde{D}$}}

\nc{\diff}{\mbox{$\frac{d}{dz}$}}
\nc{\Lder}{\mbox{$L_{-1}$}}
\nc{\bone}{\mbox{${\bf 1}$}}
\nc{\px}{\mbox{${\partial_x}$}}
\nc{\py}{\mbox{${\partial_y}$}}
\setcounter{equation}{0}

\vspace{1in}
\begin{center}
{{\LARGE\bf From String Backgrounds to Topological Field Theories}}
\end{center}
\addtocounter{footnote}{0}
\vspace{1ex}
\begin{center}
Bong H. Lian\\
Department of Mathematics\\
Brandeis University\\
Waltham, MA 02134\\
lian$@$max.math.brandeis.edu
\end{center}
\begin{center}
Gregg J. Zuckerman\\
Department of Mathematics\\
Yale University\\
New Haven, CT 06520\\
gregg$@$math.yale.edu
\end{center}
\addtocounter{footnote}{0}
\footnotetext{G.J.Z. is supported by NSF Grant DMS-9307086.}

\vspace{1ex}
\begin{quote}
{\footnotesize
Abstract:
The BRST formalism has played a fundamental role in the
construction of bosonic closed string backgrounds, ie. the stringy
analogs of classical solutions to the field equations of general
relativity. The concept of a string background has been extended
to the notion of $W$-strings, where the BRST symmetry is still
largely conjectural. More recently, the BRST formalism has
entered the construction of two dimensional topological
conformal quantum field theories, such as those that arise from
Calabi-Yau varieties.

In this lecture, we focus on common features of the BRST cohomology
algebras of string backgrounds and topological field theories. In
this context, we present some new evidence for a remarkable
relationship that transports us from bosonic and $W$-string
backgrounds to the B-model topological conformal field theories
associated to certain noncompact Calabi-Yau varieties. This
paper will appear in the proceedings of the
{\it  Symposium on  BRS Symmetry} held at RIMS, September 18-22, 1995.
}
\end{quote}

\section{Introduction}

The BRST formalism is a widely, if not universally, recognized approach
to the imposition of the Virasoro constraints in string theory
(for some early works, see \cite{KO}\cite{FMS}\cite{W1}\cite{Fe}\cite{FGZ};
see also the paper by Kato in this volume.).
Over the last dozen years physicists and mathematicians alike have
pondered the BRST-structure of string backgrounds, both abstract and
concrete.  During the
same period, conformal field theory techniques have played
an ever increasing role in the theory of string backgrounds (see
\cite{BPZ}\cite{MS}.) For a long time, the general
theory has been dominated by the standard construction of
ghost number one BRST invariant fields from dimension one
primary matter fields  (see for example \cite{FMS}).
It is well known that such standard
invariant fields form a Lie algebra,
at least modulo exact fields.

A number of research groups have understood that the operator
product expansion for the background chiral algebra leads to a
much richer algebraic structure in the {\it full} BRST cohomology,
where all ghost numbers are on an equal footing
\cite{Wi3}\cite{WZ}\cite{WuZhu}\cite{LZ9}\cite{SP}\cite{Hu}.
 In \cite{LZ9}, the
authors have recognized that the full BRST cohomology of a
conformal string background has the structure of
a Batalin-Vilkovisky algebra, which is a special type of a
Gerstenhaber algebra, or G-algebra. A G-algebra is a
generalization of a Poisson algebra, which incorporates
simultaneously the structure of a commutative algebra
and a Lie algebra \cite{Gers1}\cite{Gers2}\cite{GGS}.

The particular G-algebra of the 2D string background is especially
complex and has probably made no previous appearance in pure
mathematics. In particular, in ghost number one BRST cohomology
we obtain a (noncentral) extension of the Lie algebra of vector fields
in the plane by an infinite dimensional abelian Lie algebra \cite{LZ9}. In
this sense, string theory has enriched our understanding of
algebraic structures. At the same time, there is a connection
between the G-algebra for 2D strings and the anti-bracket
formalism. The latter appears in both the work of Schouten-Nijenhuis
\cite{nijenhuis}
on the tensor calculus as well as the work of Batalin-Vilkovisky \cite{bv}
\cite{Sch}
on the quantization of constrained field theories. Thus,
string theory has also illuminated our understanding of
geometrical structures.

As a biproduct of our structural analysis of the 2D string background,
we obtain a new and very simple description of an explicit basis for
the (chiral) BRST cohomology. This basis appears already in the work of
Witten and Zwiebach \cite{WZ}. Our enumeration of the basis exploits
the full strength of the G-algebra structure, as well as a beautiful
representation of the group $SL(2,\bC)$ in the cohomology.

In the last few years, many physicists have worked on the BRST formalism
in the context of topological conformal field theory (see the papers of
Kanno and Bonora in this volume).
It's also been shown that many conformal string backgrounds are equivalent to
TCFTs (see \cite{BLNW}). Moreover TCFTs
arise naturally in supersymmetric sigma models based on
compact Calabi-Yau manifolds.
Recently Ghoshal and Vafa have presented remarkable evidence for
the strong relationship between
certain 2D string backgrounds and TCFTs associated with
certain {\it noncompact} Calabi-Yau threefolds \cite{GoVa}\cite{OV}\cite{CO}.
This relationship provides a strong motivation for extending the theory
of sigma models to noncompact, possibly singular, varieties.
Such an extension ought to involve a prominent role for both
Gerstenhaber and Batalin-Vilkovisky algebras. Furthermore,
the mirror symmetry between the so-called A and B models in the compact case
should now be extended to the noncompact case as well as the case of
a degenerate K\"ahler structure.

In \cite{LZ16}, we raised the question of extending our results about
a 2D string background to the case of $W$-string backgrounds (see for example
\cite{BLNW}\cite{BMP}\cite{BMP2}\cite{BP}).
In the meantime, the relationship in \cite{GoVa} suggested to us a
connection between $W$-string backgrounds and certain higher dimensional
noncompact Calabi-Yau varieties \cite{TY}.

In this paper, we begin with a brief review of Gerstenhaber and
Batalin-Vilkovisky algebras. We also review the BRST formalism in the context
of chiral algebras of conformal string backgrounds.
A particular 2D string background will be
a fundamental example throughout our review. Using the
elementary notions such as the normal ordered product and the
descent operation, we show how to construct some fundamental
algebraic structures on the BRST cohomology. In the case of
the 2D string background, these structures can be described
to a large extent in terms of the geometric G-algebra of the
affine plane $\bC^2$. We also describe an explicit basis for
the BRST cohomology using the $SL(2,\bC)$ group action on
$\bC^2$. The first five sections of this paper are meant to be expository
and contain no new results beyond \cite{LZ16}.

In section \ref{sec6}, we review the notion of a topological chiral
algebra (TCA) \cite{LZ9} as well as the BV structure on the BRST cohomology
of a TCA. We then comment on a few examples of topological conformal
field theories (TCFTs). Furthermore, we discuss the A and B model TCFTs
naturally associated to the sigma model attached to a Calabi-Yau variety.
We speculate on the extension of this theory to the case of
Calabi-Yau varieties which are possibly noncompact, or possibly with
degenerate K\"ahler structure. Relevant to this discussion are the
two classical cohomology theories: the Kodaira-Spencer cohomology and
the Poisson cohomology. In section \ref{sec7}, we discuss briefly
the connection motivated by \cite{GoVa}. In section \ref{sec8},
we discuss a generalization of this connection to $W_n$-string backgrounds.
We state some conjectures on the relationship between $W_n$-string
backgrounds and certain complex varieties directly associated to
the group $SL(n,\bC)$.

\section{G-algebras and BV-algebras}\lb{sec2}

 Let $M$ be a
smooth manifold. Then the space $V^*(M)$ of polyvector fields
(ie. antisymmetric contravariant tensor fields) on
$M$ admits the structure of a G-algebra, known as the Schouten algebra of $M$.
 The dot product of a $p$-vector field $P$ and a $q$-vector $Q$ is given by
\eq
(P\cdot Q)^{\nu_1\cdots\nu_p\mu_1\cdots\mu_q}
=P^{[\nu_1\cdots\nu_p} Q^{\mu_1\cdots\mu_q]}.
\en

Now let $P,Q$ be elements of $V^{p+1}(M), V^{q+1}(M)$ respectively.
The Schouten bracket can be described as follows: in local coordinates the
Schouten bracket $[P,Q]_S$ is given by \cite{nijenhuis}
\eq
[P,Q]_S^{\nu_1\cdots\nu_p\lambda\mu_1\cdots\mu_q}=
(p+1)P^{\rho [\nu_1\cdots\nu_p}\partial_\rho Q^{\lambda\mu_1\cdots\mu_q] }
-(q+1)Q^{\rho [\mu_1\cdots\mu_q}\partial_\rho P^{\lambda\nu_1\cdots\nu_p] }.
\en
Note that if $p=q=0$, then $[P,Q]_S$ is the ordinary Lie bracket of the vector
fields $P,Q$.

\be{dfn}
A G-algebra is an integrally graded vector space $A^*$ equipped with two
bilinear products, denoted by $u\cdot v$ and $\{u,v\}$ respectively,
satisfying the following assumptions:\\
(i) If $u,v$ are homogeneous elements of degree $|u|,|v|$ respectively,
then $u\cdot v$ and $\{u,v\}$ are of degrees $|u|+|v|$ and $|u|+|v|-1$
respectively.\\
(ii) Let $u,v,t$ be elements in $A$ of degrees $|u|,|v|,|t|$ respectively. Then
we have\\
(a) $u\cdot v=(-1)^{|u||v|}v\cdot u$\\
(b) $(u\cdot v)\cdot t=u\cdot(v\cdot t)$\\
(c) $\{u,v\}=-(-1)^{(|u|-1)(|v|-1)}\{v,u\}$\\
(d)
$(-1)^{(|u|-1)(|t|-1)}\{u,\{v,t\}\}
+(-1)^{(|t|-1)(|v|-1)}\{t,\{u,v\}\}
+(-1)^{(|v|-1)(|u|-1)}\{v,\{t,u\}\}=0$\\
(e) $\{u,v\cdot t\}=\{u,v\}\cdot t +(-1)^{(|u|-1)|v|}v\cdot\{u,t\}$\\
\end{dfn}
For references on the mathematical theory of G-algebras, see
\cite{Gers1}\cite{Gers2}\cite{GGS}\cite{LZ9}(appendix B) \cite{ks}.

\be{dfn}
A BV algebra is a G-algebra, $A^*$, equipped with a linear operation $\Delta$
satisfying\\
(f) $(-1)^{|u|}\{u,v\}=\Delta(u\cdot v)-(\Delta u)\cdot v-(-1)^{|u|}u\cdot
\Delta v$\\
(g) $|\Delta|=-1$ and $\Delta^2=0$.
\end{dfn}
For references on the mathematical theory of BV algebras, see
\cite{Koszul}\cite{ks}\cite{Sch}\cite{W2}\cite{Ak}.

Let $A^*(\bC^2)$ be the holomorphic polyvector fields on $\bC^2$ with
polynomial coefficients. As a linear space, $A^*(\bC^2)$ is the super
polynomial algebra $\bC[x,y,\partial_x,\partial_y]$, where $x,y$ have degree
zero and
$\partial_x,\partial_y$ have degree one.
Introduce the notations $x^*=\partial_x$, $y^*=\partial_y$. Define the linear
operation on $A^*(\bC^2)$: $D=\frac{\partial}{\partial x}
\frac{\partial}{\partial x^*} +\frac{\partial}{\partial y}
\frac{\partial}{\partial y^*}$.
\be{pro}
$A^*(\bC^2)$ equipped with $D$ is a BV algebra.\\
\end{pro}
(See \cite{LZ9}\cite{W2}.)

\section{BRST Formalism for String Backgrounds}\lb{sec3}

The chiral algebra of a conformal string background is of the form
\eq
\cA=\cA^{ghost}\otimes\cA^{matter}
\en
where $\cA^{ghost}$, $\cA^{matter}$ are respectively the chiral algebras
of the ghost and the matter sectors. Thus explicitly an element $\cO$ of $\cA$
is a finite
sum of holomorphic fields of the form $P(b,\partial b,\cdots,c,\partial
c,\cdots) \Phi$, where $P$ is a normal ordered differential
polynomial of the ghost fields
$b,c$; $\Phi$ is a field in the matter sector. The respective (holomorphic
part of the) stress energy tensors of the two sectors are
$L^{ghost},L^{matter}$
whose central charges are respectively $c=26$, and $c=-26$. The BRST current is
\eq
J=c(L^{matter}+\half L^{ghost}),
\en
and the ghost number current is
\eq
F=cb.
\en

Let's recall a few basics of the BRST formalism. The BRST charge
\eq
Q=\oint_{C_0}J(z)dz
\en
has the property that $[Q,Q]=2Q^2=0$, where square brackets $[\cdot,\cdot]$
denotes the super commutator. Here $C_0$ is a small contour around $0$.
A BRST invariant field $\cO$ is one which satisfies
$[Q,\cO]=0$; and a BRST exact field is one which is
 of the form $\cO=[Q,\cO']$ for some
field $\cO'$. The BRST cohomology of the string background $\cA$ is
the quotient space
\eq
H^*(\cA)=\mbox{$\{Q$-invariant fields$\}/\{Q$-exact fields$\}$}
\en
which is graded by the ghost number $*$.

A standard way
to obtain BRST invariants is as follows.
Let $V$ be a primary field of dimension one in $\cA^{matter}$. Then
both $cV$ and $c\partial cV$ are BRST invariants of ghost number 1,2
respectively. The fields $\cO=1$ and $\cO=c\partial c\partial^2 c$ are two
universal BRST invariants known from the bosonic string theory.

\subsection{An example of a string background}\lb{3.1}

The chiral algebra $\cA_{2D}$ of a particular 2D string background is generated
by the fields $b,c,\partial X,\partial\phi$, $e^{\pm(\pm iX-\phi)/\sqrt{2}}$,
where $X,\phi$ are free bosons with the usual OPEs. Their corresponding
stress energy tensors are:
\eqa
L^X&=&-\half(\partial X)^2\nnb\\
L^\phi&=&-\half(\partial\phi)^2+\sqrt{2}\partial^2\phi
\ena
whose respective central charges are $c=1, c=25$.

We now describe some BRST invariants of the  chiral algebra $\cA_{2D}$.
Since $e^{(\pm iX+\phi)/\sqrt{2}}$ are matter primary fields of dimension one,
we immediately obtain some standard BRST invariants $-ce^{(\pm
iX+\phi)/\sqrt{2}}$. We denote them by $Y^+_{1/2,\pm1/2}$. There are
also {\it exotic}
BRST invariants which cannot be obtained in the above standard way. For
example, in ghost number zero we have
\eq
\cO_{1/2,\pm1/2}=\left(cb+\frac{i}{\sqrt{2}}(\pm\partial
X-i\partial\phi)\right)
e^{(\pm iX-\phi)/\sqrt{2}}.
\en
These BRST invariants were identified in \cite{LZ4} (see also \cite{BMP}),
and their explicit formulas were given in \cite{Wi3}\cite{WZ}. Explicit
formulas for infinitely many BRST invariants of the 2D string background
are also known. See \cite{BMP}\cite{MMS}\cite{WuZhu}\cite{WZ}.

\section{Fundamental Operations}\lb{sec4}

Using the antighost field $b$ and contour integral, one can define a number
of useful operations on the fields $\cO$ in a string background. Given
a dimension $h$ field $\cO$,
we attach to it a field of dimension $h+1$
\eq
\cO^{(1)}(w)=\oint_{C_w}b(z)\cO(w)dz
\en
where $C_w$ is a small coutour around $w$.
We call this linear operation, which reduces ghost number by one, the descent
operation. Similarly, to $\cO$ we can attach a field
of dimension $h$
\eq
\Delta \cO(w)=\oint_{C_w}b(z)\cO(w)(z-w)dz.
\en
This linear operation is called the Delta operation.

There are two important bilinear operations defined as follows. The first
one is the dot product (a.k.a. the normal ordered product). Given two fields
$\cO_1,\cO_2$, we define
\eq
\cO_1(w)\cdot\cO_2(w)=\oint_{C_w}\cO_1(z)\cO_2(w)(z-w)^{-1}dz.
\en
Under the dot product, both the conformal dimension and the ghost number are
additive.

We define also the bracket operation
\eq
\{\cO_1(w),\cO_2(w)\}=\oint_{C_w}\cO_1^{(1)}(z)\cO_2(w)dz.
\en
Note that conformal dimension is additive, while ghost number is
shifted by -1 under the bracket operation. This operation was implicit
in \cite{Wi3}\cite{WZ}, and was studied in general in \cite{LZ9}.

Let
\eqa
Q*\cO(w)&=&\oint_{C(w)}J(z)\cO(w)dz=[Q,\cO(w)]\nnb\\
\Sigma*\cO(w)&=&\oint_{C(w)}L^{total}(z)\cO(w)(z-w)dz.
\ena
Then we have the following algebra of operations:
\eqa\lb{3.13}
\partial\cO&=&{[Q,\cO^{(1)}]}\nnb\\
\Delta^2&=&0\nnb\\
(Q*)^2&=&0\nnb\\
{[Q*,\Delta]}&=&\Sigma*\nnb\\
\Sigma*\cO&=&n\cO \mbox{ iff  the conformal dimension of $\cO$ is $n$.}
\ena
Moreover, we have the following identities \cite{LZ9}:
\eqa\lb{3.14}
&(a)& Q*(\cO_1\cdot\cO_2)=(Q*\cO_1)\cdot\cO_2+(-1)^{|\cO_1|}\cO_1\cdot
(Q*\cO_2)\nnb\\
&(b)& Q*\{\cO_1,\cO_2\}= \{Q*\cO_1,\cO_2\}+(-1)^{|\cO_1|-1}
\{\cO_1,Q*\cO_2\}\nnb\\
&(c)& (-1)^{|\cO_1|}\{\cO_1,\cO_2\}=
\Delta(\cO_1\cdot\cO_2)-(\Delta\cO_1)\cdot\cO_2
-(-1)^{|\cO_1|}\cO_1\cdot\Delta\cO_2\nnb\\
\ena
where $|\cO|$ denotes the ghost number of $\cO$.
Note that the first equation in \erf{3.13} is known as the
descent equation \cite{WZ}.

\subsection{Induced algebraic structures in BRST cohomology}

Let $[\cO], [\cO_1], [\cO_2]$ be cohomology classes in $H^*(\cA)$.
By virtue of the identities \erf{3.13} and \erf{3.14}, the above
operations induce the following well-defined operations on cohomology:
\eqa
\Delta [\cO]&=&[\Delta\cO_{[0]}]\nnb\\
{[\cO_1]}\cdot{[\cO_2]}&=&{[\cO_1\cdot\cO_2]}\nnb\\
\{[\cO_1],[\cO_2]\}&=& {[\{\cO_1,\cO_2\}]}.
\ena
Here $\cO_{[0]}$ is the projection of $\cO$ onto the subspace of $\cA$
consisting of fields of zero conformal dimension.
For example, given the standard BRST cohomology classes $\cO_i=cV_i$, $i=1,2$,
where the $V_i$ are matter primary fields of conformal dimension one, we have
\eq
\{[\cO_1],[\cO_2]\}=[\cO_3]
\en
where
\eq
\cO_3(w)=c(w)\oint_{C_w}V_1(z)V_2(w)dz.
\en

\subsection{Chiral ground ring}

Since ghost number is additive under dot product in cohomology, it follows
that $H^0(\cA)$ is closed under this product.

\be{thm}\cite{Wi3}
The dot product in $H^0(\cA)$ is commutative and associative.
\end{thm}
The commutative associative algebra $H^0(\cA)$ is called the chiral
ground ring of $\cA$.
For example it is shown in \cite{Wi3}\cite{LZ9} that in the case
$\cA=\cA_{2D}$, the ground ring is the polynomial algebra generated by
the classes ${[\cO_{1/2,1/2}]}$ and ${[\cO_{1/2,-1/2}]}$.

\subsection{Chiral cohomology ring}

\be{thm}\cite{LZ9}
$H^*(\cA)$ is a BV-algebra with BV operator $\Delta$ and with the dot and
bracket products defined above.
\end{thm}
For related papers on algebraic structure in BRST cohomology, see
\cite{FGZ}\cite{GJ}\cite{Hu}\cite{KSV}\cite{KSV2}\cite{LZ10}
\cite{M}\cite{SP}\cite{WuZhu}\cite{Ak}.

\section{A 2D String Background}\lb{sec5}

An important example of a G-algebra is the one given by the BRST cohomology of
the 2D string background discussed in section \ref{3.1}:
\be{thm}
As a G-algebra, $H^*(\cA_{2D})$ is generated by the four classes
${[\cO_{1/2,\pm 1/2}]}$, ${[Y^+_{1/2,\pm 1/2}]}$. Moreover, $H^p(\cA_{2D})$
vanishes except for $p=0,1,2,3$.
\end{thm}
Note that $\{ {[Y^+_{1/2,- 1/2}]}, {[Y^+_{1/2, 1/2}]} \}=
{[ce^{\sqrt{2}\phi}]}\neq0$. (See \cite{LZ9}.)

\be{thm}
The assignment ${[\cO_{1/2, 1/2}]}\mapsto x$,  ${[\cO_{1/2, -1/2}]}\mapsto y$,
${[Y^+_{1/2, -1/2}]}\mapsto \partial_x$, ${[Y^+_{1/2, 1/2}]}\mapsto \partial_y$
extends to a G-algebra
homomorphism $\psi$ of $H^*(\cA_{2D})$ onto $A^*(\bC^2)$.
\end{thm}
For details on this result, see \cite{LZ9}.

\be{pro}
For $u$ in $H^*(\cA_{2D})$, $\psi(\Delta u)=- D(\psi u)$.
\end{pro}
(See \cite{LZ9}\cite{WZ}.)

Introduce the notations $\tilde{x}={[\cO_{1/2, 1/2}]}$,  $\tilde{y}={[\cO_{1/2,
-1/2}]}$,
$\tilde{\partial_x}={[Y^+_{1/2, -1/2}]}$, $\tilde{\partial_y}={[Y^+_{1/2,
1/2}]}$,
 $\tilde{J}_+=\tilde{x}\tilde{\partial_y}$,
$\tilde{J}_0=\tilde{x}\tilde{\partial_x}-\tilde{y}\tilde{\partial_y}$,
$\tilde{J}_-=\tilde{y}\tilde{\partial_x}$. Similarly let
$J_+,J_0,J_-\in A^*(\bC^2)$ be
defined by analogous formulas but without the tildes.
\be{pro}
$Span\{\tilde{J}_+, \tilde{J}_0, \tilde{J}_-\}$,
$Span\{J_+,  J_0, J_-\}$ are both closed under the bracket $\{\cdot,\cdot\}$
and are isomorphic to the Lie
algebra $sl(2,\bC)$.
\end{pro}
(See \cite{Wi3}\cite{WZ}\cite{LZ9}\cite{BMP}.)

Now introduce an $sl(2,\bC)$-action on $H^*(\cA_{2D})$ by
$\tilde{J}_a*\cO=\{\tilde{J}_a, \cO\}$ where $a=+,0,-$, $\cO\in H^*(\cA_{2D})$.
Similarly, introduce an $sl(2,\bC)$-action on $A^*(\bC^2)$
 by $J_a*X=\{J_a, X\}$, $X\in A^*(\bC^2)$.
We make the observation (see \cite{LZ9}\cite{BMP}) that
$\psi:H^*(\cA_{2D})\ra A^*(\bC^2)$ intertwines actions of $sl(2,\bC)$.

\be{pro} \cite{LZ9}
$ker\ \psi$ is a $G$-ideal with vanishing dot product and bracket product.
\end{pro}
That is, for $\cO\in H^*(\cA_{2D})$, $\cO',\cO''\in ker\ \psi$,
we have $\cO\cdot\cO', \{\cO,\cO'\}\in ker\ \psi$, and
$\cO'\cdot\cO''=\{\cO',\cO''\}=0$.

\be{pro} \cite{LZ9}
$ker\ \psi$ as a $G$-ideal is generated by the class ${[ce^{\sqrt{2}\phi}]}$.
\end{pro}
This means that ${[ce^{\sqrt{2}\phi}]}\in ker\ \psi$, and the smallest subspace
containing ${[ce^{\sqrt{2}\phi}]}$ and
stable under the action of $H^*(\cA_{2D})$ by both the
dot product and the bracket product is the whole $ker\ \psi$ itself.

Let's give an explicit basis for $ker\ \psi$. Introduce the following linear
operations on cohomology classes $[\cO]$:
\eqa
A[\cO]&=&\{\tilde{\partial_x},[\cO]\}\nnb\\
B[\cO]&=&\{\tilde{\partial_y},[\cO]\}\nnb\\
C[\cO]&=&\tilde{\partial_x}\cdot[\cO]\nnb\\
D[\cO]&=&\tilde{\partial_y}\cdot[\cO].
\ena
Fix $\cK=ce^{\sqrt{2}\phi}$. Then we have
\be{pro}
$ker\ \psi$ as a vector space has a basis consisting of the following classes:
\eqa
&ghost\ no.\ 1:& A^{s-n}B^{s+n}[\cK]\nnb\\
&ghost\ no.\ 2:& (s-n)A^{s-n-1}B^{s+n}C[\cK]+
(s+n)A^{s-n}B^{s+n-1}D[\cK]\nnb\\
&ghost\ no.\ 2:& A^{s-n+1}B^{s+n}D[\cK]-
A^{s-n}B^{s+n+1}C[\cK]\nnb\\
&ghost\ no.\ 3:& A^{s-n}B^{s+n}CD[\cK]
\ena
where $s=0,\half,1,...,$ and $n=-s,-s+1,..,s$.
\end{pro}
We observe that the above basis vectors are in fact weight vectors for
the $sl(2,\bC)$ action which we have previously described. Here $s,n$
are respectively the total spin and axial spin quantum numbers of the
$sl(2,\bC)$ representation.

For completeness, we list here a basis for the algebra
$A^*(\bC^2)\cong H^*(\cA_{2D})/ker\ \psi$:
\be{eqnarray}
&ghost\ no.\ 0:&x^{s-n}\cdot y^{s+n}\nnb\\
&ghost\ no.\ 1:&\px(x^{s-n}\cdot y^{s+n})\cdot\py
-\py(x^{s-n}\cdot y^{s+n})\cdot\px\nnb\\
&ghost\ no.\ 1:&x^{s-n}\cdot y^{s+n}\cdot(x\cdot\px+y\cdot\py)\nnb\\
&ghost\ no.\ 2:&x^{s-n}\cdot y^{s+n}\cdot\px\cdot\py
\end{eqnarray}
where $s=0,\half,1,...,$ and $n=-s,-s+1,..,s$.

\section{BRST Formalism for Topological Conformal Field Theories}\lb{sec6}

\be{dfn}\cite{LZ9}
A topological chiral algebra (TCA) consists of the following data: a chiral
(super)algebra $C^*$, a weight one even current $F(z)$ whose charge
$F_0$ is the fermion number operator, a weight one odd primary field
$J(z)$ having fermion number one and having a square zero charge $Q$, and
a weight two odd primary field $G(z)$ having fermion number -1 and
satisfying $[Q,G(z)]=L(z)$, where $L(z)$ is the stress-energy field.
We denote the cohomology of the complex $(C^*,Q)$ by $H^*(C)$.
We call the charge $Q$ the BRST operator of the TCA $C^*$.
\end{dfn}

\be{thm}\cite{LZ9}
Given a TCA $C^*$, $H^*(C)$ is a BV-algebra with BV operator $G_0$ and
with dot product induced by the Wick product.
\end{thm}

	Given a particular conformal field theory \cite{BPZ}, we can begin with the
full operator algebra and then extract the chiral algebra \cite{MS} as the
subalgebra of purely holomorphic operators.  Since the spin of any operator in
the full operator algebra is an integer, the dimension of any holomorphic
operator is again an integer.   In the same say, we can begin with the full
operator (super)algebra of a topological conformal field theory--TCFT--and
extract the topological chiral algebra--TCA--of purely holomorphic operators
\cite{DVV}\cite{BCOV}.  Just as a TCA has a cohomology ring which is a BV
algebra, the full TCFT operator algebra has a cohomology ring with is also a BV
algebra\cite{GJ}\cite{SP}\cite{KSV},
and which contains the cohomology ring of the TCA.

For example, let
\eq
\cB=\cA_{2D}\otimes\bar{\cA}_{2D}
\en
where $\bar{\cA}_{2D}$ is the antiholomorphic counterpart of $\cA_{2D}$.
We can regard $\cB$ as the ``bilateral'' operator algebra of a particular
TCFT with the BRST operator $Q+\bar{Q}$. The associated TCA is just $\cA_{2D}$
itself. The BRST cohomology of $\cB$ is given by
$H^*(\cA_{2D})\otimes H^*(\bar{\cA}_{2D})$ as a BV-algebra with the
BV operator $\Delta+\bar{\Delta}$.

For a more interesting example, we can take $\cB'$ be the subalgebra of
$\cB$ consisting of operators with zero charge relative to current
$\partial\phi-\bar{\partial}\phi$. In this case $Q+\bar{Q}$ is still
the BRST operator, but the cohomology is a proper subalgebra of
$H^*(\cA_{2D})\otimes H^*(\bar{\cA}_{2D})$. The TCA associated with $\cB'$ is a
subTCA of $\cA_{2D}$ generated by the holomorphic operators
$b,c, e^{\pm i\sqrt{2}X}, \partial \phi$. The ground ring \cite{Wi3}
of the string
background $\cB'$ is the degree zero BRST cohomology of $\cB'$. This is
a commutative algebra generated by
\eqa
x_1&=&[\cO_{1/2,1/2}][\bar{\cO}_{1/2,1/2}]\nnb\\
x_2&=&[\cO_{1/2,1/2}][\bar{\cO}_{1/2,-1/2}]\nnb\\
x_3&=&[\cO_{1/2,-1/2}][\bar{\cO}_{1/2,1/2}]\nnb\\
x_4&=&[\cO_{-1/2,-1/2}][\bar{\cO}_{-1/2,-1/2}],
\ena
with one relation: $x_4x_1-x_2x_3=0$. In other words,
\eq
H^0(\cB')\cong \bC[x_1,..,x_4]/(x_4x_1-x_2x_3)
\en
the right hand side being the coordinate ring of the quadric cone in $\bC^4$.

\subsection{TCFTs and Calabi-Yau varieties}

	Because a fermion number current $F(z)$ appears as a special element of a TCA,
a TCFT will have both a bosonic and fermionic sector, unlike a typical CFT,
which is purely bosonic.  Thus, a TCFT is a cousin of a supersymmetric
conformal field theory.  In fact, there is a
standard construction that in two ways twists an N = 2 super CFT to produce
two TCFTs, the so-called A and B models.  Moreover, many TCFTs are known to
arise as twisted N = 2 theories\cite{DVV}.
	In the same way, many TCAs arise from twisting an N = 2 chiral superalgebra.
The TCA $A_{2D}$ analyzed previously
 is a good example of a twisted N = 2 chiral superalgebra.  Vafa and Mukhi
\cite{MVa} have made a study of this particular twisting.

In the following, we will discuss the recent and exciting interaction between
the theory of TCFTs and the theory of Calabi-Yau varieties.

1) Cohomology rings of mirror manifolds (See for example \cite{Va} \cite{W4}
\cite{BCOV}):  Physicists hypothesize the existence of an N = 2 supersymmetric
CFT ("the supersymmetric sigma model") associated to a given CY variety.
Although the full state space and operator algebra of this N = 2 theory are not
in general describable in closed form, the so-called topological sectors
of the A and B twisted theories are at least partially understood.  In
particular, the A and B model cohomology rings have been identified with
possibly deformed versions of classical cohomology rings of the CY variety.   A
remarkable consequence of this identification is the famous mirror variety
hypothesis.  We describe below some features of the A and B models:

The A-model:  For a smooth, compact and K\"ahler CY variety, the cohomology
ring of  the A model is in general a nontrivial ``quantum'' deformation of its
classical counterpart, the deRham cohomology.   The only natural BV operator in
de Rham cohomology is the zero operator.   In fact, the natural BV operator on
the cohomology of the A-model TCFT is also the zero operator.

The B-model:  The B model cohomology ring is precisely isomorphic to its
classical counterpart, the Kodaira-Spencer cohomology ring:
\be{dfn} Let X be a complex manifold, let $\cO$ be the sheaf of holomorphic
functions, and let $\Theta$ be the sheaf of holomorphic vector fields.  The
Kodaira-Spencer cohomology ring \cite{BCOV} is the sheaf cohomology
$H^*(X, \wedge^*_{\cal O}\Theta)$.
\end{dfn}

One can regard $H^*(X, \wedge^*_{\cal O}\Theta)$ as the Dolbeault
$\bar{\partial}$-cohomology
of $X$ with coefficients in holomorphic polyvector fields.

If the $n$-dimensional complex manifold $X$ has trivial canonical bundle, the
holomorphic volume form leads to an isomorphism of graded vector spaces:
\eq
H^*(X,\wedge^*_{\cal O}\Theta)\ra H^*(X,\wedge^{n-*}_{\cal O}\Omega)
\en
Here $\Omega$ is the sheaf of holomorphic 1-forms.
Thus $H^*(X,\wedge^{*}_{\cal O}\Omega)$ can be regarded as the
$\bar\partial$-cohomology of $X$ with coefficients in holomorphic forms.
 The operator
$\partial$ on $\Omega$ induces an operator on the range of the above
isomorphism, and in turn a natural BV operator $\Delta$ on the Kodaira-Spencer
ring itself.  If $X$ is a smooth, compact and K\"ahler CY variety,
then $\Delta$ vanishes identically.

{\it Comparison of the models}:  Recall that for a smooth projective CY
variety, only the de Rham cohomology is quantum-deformed.  This observation is
of course the key to the successful applications of the mirror symmetry
hypothesis.   However, one can ask whether this asymmetry between the A and B
model continues to hold for more general complex manifolds.

2) Generalizations of cohomology rings: We also see from the above that for a
smooth projective CY variety, both of the classical cohomology rings have
trivial BV structure.  At the level of topological conformal field theory,  the
respective BV operators are likewise trivial in both the A and B model
cohomology rings attached to $X$.  Thus, only the graded commutative (dot)
product is nontrivial; the graded Lie bracket product vanishes identically.
Thus there are two distinguished asymmetries: one between the A and the B
models of $X$, and one between the dot and the bracket products of the BV
algebras associated to $X$.
We have been puzzled by these asymmetries in the case of
 smooth projective CY varieties.  We have therefore been studying varieties of
a more general type. It will be easier to consider the B model first.

{}From the purely mathematical definition of the Kodaira-Spencer cohomology, we
have observed the following:

i) For any complex manifold, there is a natural bracket product which makes
 the Kodaira-Spencer cohomology a $G$-algebra. This bracket is induced
by the Schouten bracket on holomorphic polyvector fields.

ii) For a complex manifold with trivial canonical bundle (a ``zero canonical
manifold''), the natural BV operator on the Kodaira-Spencer ring induces the
natural bracket in i).

iii) For noncompact complex manifolds, the natural bracket can be nontrivial;
likewise for noncompact zero canonical manifolds, the BV operator can be
nontrivial.  Simple examples arise as open dense submanifolds of compact
complex manifolds (see the next section).

In principle, the physicists' theory of smooth
projective CY varieties ought to extend to some class of noncompact complex
varieties.  In particular, such an extension should include the open subvariety
of smooth points (the "smooth locus") on a singular CY variety.  In this way,
the BV-structure of TCFT theory should come to play an important  role.  We
therefore propose to continue our study of the Kodaira-Spencer cohomology of
complex varieties.

{\it The A model analog}:   Unfortunately, the de Rham cohomology of a smooth
manifold (compact or noncompact)
has no natural nontrivial BV algebra structure. However, the A model  involves
in an essential way the space of K\"ahler forms on the CY variety.  A K\"ahler
form gives rise to a Poisson tensor, which is of course nondegenerate, and
completely determines the K\"ahler form itself.

\be{dfn}
Let $Y$ be a smooth manifold and let $V^*(Y)$
be the smooth sections of the Grassman algebra bundle generated by the tangent
bundle of $Y$. $V^*(Y)$ is a G-algebra relative to the wedge and Schouten
products.  Let $P$ in $V^2(Y)$ be a Poisson bivector field on $Y$.
We call $P$ a Poisson tensor if the Schouten bracket $[P,P]=0$.
\end{dfn}

It is well-known that the inverse of a K\"ahler (in fact even symplectic)
form is a Poisson tensor.
For arbitrary Poisson tensors, there is a geometrical invariant known as the
Poisson cohomology, which was introduced by Lichnerowicz \cite{Li} and later
studied by Koszul \cite{Koszul}. For more details, see these two references.

\be{dfn}\cite{Li}
The operator $\delta_P=[P,-]$ is square zero and is a derivation of degree one
of the G-algebra structure on $V^*(Y)$.  Finally, the Poisson cohomology ring
of Y, $H^*_P(Y)$, is the cohomology of $\delta_P$ in $V^*(Y)$.
\end{dfn}

If the Poisson tensor is nondegenerate (invertible), the Poisson cohomology is
naturally isomorphic to the de Rham cohomology.  The Poisson cohomology is, for
any Poisson tensor, a Gerstenhaber algebra; in some circumstances it is even a
BV algebra.   We have found simple examples with nontrivial G-algebra or
BV-algebra structure.
We propose to study a generalization of the N = 2 supersymmetric sigma model to
the case where the K\"ahler form is replaced by a degenerate Poisson tensor
\cite{SS}.  For such target spaces, the A model cohomology should in some cases
be a nontrivial BV algebra.

3) Degenerations of complex and K\"ahler structure:  We hope to eventually
generalize the mirror variety hypothesis beyond the context of smooth
projective CY varieties.   The new context should at least include Poisson
varieties that arise by degeneration of K\"ahler structure, as well as the
smooth loci of singular varieties that arise by degeneration of the complex
structure.  We have found simple examples to illustrate the notion of a
degeneration of K\"ahler structure on a complex manifold.  Of course,
degeneration of complex structure is a well studied topic in the theory of
Calabi-Yau varieties (see for example \cite{CO}).

	 The generalized mirror variety hypothesis should include a description of how
the BV algebra structure of the A model produces a quantum
deformation of the BV algebra structure of the Poisson cohomology.  Likewise,
there must be a description of how the B-model BV algebra produces a
quantum deformation of the Kodaira-Spencer cohomology.

\section{From The 2D String to Quadric Threefolds}\lb{sec7}

	We have discovered a potential example of a B-model deformation:

	Consider the quadric cone in ${\bf C}^4$.  This cone has a nodal singularity
at the vertex.  Suppose that $X$ is the cone minus the vertex.  $X$ is a zero
canonical manifold, and we have calculated its Kodaira-Spencer cohomology as a
BV algebra.  Following a striking suggestion by Ghoshal and Vafa \cite{GoVa},
we have compared the Kodaira-Spencer cohomology of $X$ with the bilateral BRST
cohomology of the TCFT $\cB'$, which arises
in the theory the $c=1$ string (see \ref{sec6}) \cite{Wi3} \cite{WZ}
\cite{LZ4}.  The two cohomologies are nearly isomorphic as graded commutative
algebras, but the G-bracket and BV operator of the c = 1 string TCFT are
nontrivial deformations of the corresponding bracket and operator on the
Kodaira-Spencer cohomology.

	We expect that there is a generalization of the B-model to
	the case of a noncompact zero canonical variety (such as the cone $X$). We
conjecture that
the B-model cohomology of the cone $X$ is isomorphic to the BRST cohomology of
the c = 1 string TCFT.  In principle, there should even be an equivalence at
the level of TCFTs.   There is remarkable evidence in the Ghoshal-Vafa paper
\cite{GoVa} for an equivalence between the TCFT of the smooth quadric and the
TCFT obtained by perturbing the c = 1 string theory by an appropriate marginal
operator \cite{Wi3}.   Whether or not this evidence is relevant to the
singular cone is still not clear.

	We propose to continue our study of particular complex varieties which may be
noncompact rather than projective and/or Poisson rather than K\"ahler.  Our
main goal is the precise formulation of a generalized mirror symmetry
hypothesis, as well as the application of such a hypothesis to some specific
families of CY varieties.

\section{From the $W_n$-String Background to a Degeneration of $SL(n,{\bf
C})$}\lb{sec8}

The c = 1 string TCFT is the first in a series of TCFTs called W-string
backgrounds \cite{BLNW}.  Such backgrounds are associated with simply-laced
simple Lie algebras over the complex numbers.  The c = 1 string theory itself
is connected with $sl(2, \bC)$.   The so-called
$W_n$-string background is connected with $sl(n, \bC)$.

\be{con}
   For arbitrary $n > 1$, there is a canonical BRST current for a $W_n$-string
background.  The cohomology of the corresponding TCA admits a natural
filtration that respects the natural BV algebra structure.  The associated
graded BV algebra is isomorphic to the Kodaira-Spencer BV algebra of the
noncompact complex variety (the ``base affine space'')
\[
X_n=SL(n,\bC)/R_n
\]
where $R_n$ is the group of $n \times n$ upper triangular unipotent matrices.
This isomorphism respects the natural action of $SL(n, \bC)$
on the $W_n$ BV algebra and the Kodaira-Spencer BV algebra of $X_n$,
respectively.
\end{con}

A critical step in the development of the above conjecture was an
earlier conjecture by Bouwknegt, McCarthy and Pilch: the algebra of
holomorphic polyvector fields on the base affine space is isomorphic
as a graded commutative algebra to a distinguished subalgebra of the
chiral cohomology of the $W_n$-string background \cite{BLNW}\cite{BMP2}
\cite{BP}\cite{BP1}\cite{Mc}. This statement has been verified
for the case of $n=2$ \cite{LZ9} and $n=3$ \cite{BMP2}.

Let $D_n$ be the group of diagonal matrices in $SL(n,{\bf C})$. Define
an action of $D_n$ on the Cartesian product $X_n\times X_n$ by the
formula
\eq
h\cdot(gR_n,g'R_n)=(ghR_n,g'h^{-1}R_n).
\en
Let $Z_n$ be the quotient of $X_n\times X_n$ by the above action of $D_n$;
this action is free and $Z_n$ is a smooth zero canonical noncompact complex
manifold. In fact, $Z_n$ is the smooth locus of a particular algebraic
degeneration of the group variety, $SL(n,{\bf C})$. For $n=2$ we obtain
the quadric cone discussed in the previous section. By a general theorem
of Tian and Yau \cite{TY}, $Z_n$ supports Calabi-Yau metrics.

For each $n$ there is an appropriate TCFT $\cB_n'$ generalizing the case
of $\cB_2':=\cB'$ (see previous section).
\be{con}
   For arbitrary $n > 1$,  the cohomology of the corresponding TCFT
$\cB_n'$ is nearly isomorphic as a graded commutative algebra to
the Kodaira-Spencer algebra of
the CY variety $Z_n$.
\end{con}

We propose to compute by Kodaira-Spencer cohomology of the
homogeneous complex manifold $X_n$
by means of a standard reduction \cite{Ha} to a Lie algebra cohomology of
the Lie algebra of  $R_n$. A similar reduction applies to the homogeneous
 complex manifold $Z_n$. We have recently learned that
 Bouwknegt, McCarthy and Pilch have also observed that the $R_n$-
 Lie algebra cohomology theory plays a crucial role in the cohomology
 theory of $W_n$-string backgrounds. See reference
 \cite{BP} for an early hint of this role. We look forward to future
 progress on the above conjectures.

{\bf Acknowledgements.} G.J.Z. thanks the organizers of the Symposium on the
BRS Symmetry --
special thanks to N. Nakanishi -- for their invitation to lecture.
We also thank P. Bouwknegt, B. Khesin, J. McCarthy, G. Moore, K. Pilch,
A. Todorov, C. Vafa and S.T. Yau for useful discussions.


\begin{thebibliography}{99} \spacingset{1.0}

\bibitem{Ak} F. Akman, ``On Some Generalizations of Batalin-Vilkovisky
Algebras'', q-alg/9506027.

\bibitem{bv}
I.A. Batalin and G.A. Vilkovisky, ``Quantization of gauge theories with
linearly dependent generators'', Phys. Rev. D28 (1983) 2567.

\bibitem{BPZ}
A. Belavin, A.M. Polyakov and A.A. Zamolodchikov,
``Infinite conformal symmetry in two dimensional quantum field theory'',
Nucl. Phys. B241 (1984) 333.

\bibitem{Bor}
R.E. Borcherds,
``Vertex operator algebras, Kac-Moody algebras and the Monster'',
Proc. Natl. Acad. Sci. USA. 83 (1986) 3068.

%
\bibitem{BMP}
P. Bouwknegt, J. McCarthy and K. Pilch, ``BRST analysis of physical
states for 2D gravity coupled to $c\leq1$ matter'', Commun. Math. Phys.
145 (1992) 541.

\bibitem{BMP2}
P. Bouwknegt, J. McCarthy and K. Pilch,   ``The $W_N$ algebra:  modules,
semi-infinite cohomology and BV algebras," hep-th/ 9509119.

\bibitem{BP1}
P. Bouwknegt, K. Pilch, ``The BV-algebra structure of $W_3$ cohomology'',
preprint 1994, to appear in the Proceedings of ``G\"ursey Memorial Conference
I: Strings and Symmetries'', eds. M. Serdaroglu et al.

\bibitem{BP}
P. Bouwknegt, J. McCarthy and  K. Pilch,
``On the W-gravity spectrum and its G-structure'',
in the proceedings of the workshop  ''Strings, Conformal Models and
Topological Field Theory'', eds. L. Baulieu et al. , pp59-70,
Cargese, May '93, Plenum Press, New York, 1995. hep-th/9311137.

\bibitem{BLNW} M. Bershadsky, W. Lerche, D. Nemeschansky and N. Warner,
``Extended N = 2 superconformal structure of gravity and W-gravity coupled to
matter," Nucl. Phys. B401 (1993) 304-.

\bibitem{BCOV} M. Bershadsky, S. Cecotti, H. Ooguri and C. Vafa,
``Kodaira-Spencer theory of gravity and exact results for quantum string
amplitudes," Comm. Math. Phys. 165 (1994) 311-.

\bibitem{CO} P. Candelas and X.C. de la Ossa,   ``Comments on Conifolds," Nucl.
Phys. B342 (1990) 246-268.

\bibitem{DVV} R. Dijkgraaf, E. Verlinde and H. Verlinde,  ``Notes on
topological string theory and 2D quantum gravity," IASSNS-HEP-90/80.

\bibitem{Fe}
 B. Feigin,  ``Semi-infinite homology of the Kac-Moody and Virasoro algebras,"
Russian Math. Surveys  39 (1984), No. 2, 155.

%
%
%
\bibitem{FGZ}
I.B. Frenkel, H. Garland and G.J. Zuckerman,
``Semi-infinite cohomology and string theory'',
Proc. Nat. Acad. Sci. U.S.A. 83 (1986) 8442.

%
\bibitem{FLM}
I.B. Frenkel, J. Lepowsky and A. Meurman,
``Vertex Operator Algebras and the Monster'',
Academic Press, New York, 1988.

\bibitem{FMS}
D. Friedan, E. Martinec and S. Shenker,
``Conformal invariance, supersymmetry and string theory'',
Nucl. Phys. B271 (1986) 93.

\bibitem{Gers1}
M. Gerstenhaber,
``The cohomology structure of an associative ring'',
Ann. of Math. 78, No.2 (1962) 267.

\bibitem{Gers2}
M. Gerstenhaber,
``On the deformation of rings and algebras'',
Ann. of Math. 79, No.1 (1964) 59.

\bibitem{GGS}
M. Gerstenhaber, A. Giaquinto and S. Schack,
``Quantum symmetry'', Univ of Penn preprint 1991.

\bibitem{GJ}
E. Getzler, ``Batalin-Vilkovisky algebras and two-dimensional topological field
theory," Comm. Math. Phys. 159 (1994) 265-285.

\bibitem{GoVa} D. Ghoshal and C. Vafa, ``c=1 string as the topological theory
of the conifold," hep-th/ 9506122.


\bibitem{Ha} W. Haboush, ``Homogeneous vector bundles and reductive subgroups
of reductive algebraic groups," American Journal of Mathematics   Vol. 100 No.
6 (1978) 1123-1137.

\bibitem{Hu}
Y-Z. Huang, ``Operadic formulation of topological vertex algebras and
Gerstenhaber or Batalin-Vilkovisky algebras'', Univ of Penn preprint (1993).

%
\bibitem{KO}
M. Kato and K. Ogawa,
``Covariant quantization of string based on BRS invariance'',
Nucl. Phys. B212 (1983) 443.

\bibitem{KSV}
T. Kimura, J. Stasheff and A. Voronov,
``On operad structures of moduli spaces and string theory'',
hep-th/9307114.

\bibitem{KSV2}
T. Kimura, J. Stasheff and A. Voronov,
``Homology of moduli spaces of curves and commutative homotopy algebras'',
alg-geom/9502006.

\bibitem{ks}
Y. Kosmann-Schwarzbach, ``Exact Gerstenhaber algebras and Lie bialgebroids'',
U.R.A. 169 CNRS preprint 1994.

\bibitem{Koszul}
J.-L. Koszul, ``Crochet de Schouten-Nijenhuis et cohomologie'',
Asterique, hors serie, (1985) 257-271.

\bibitem{Li}
A. Lichnerowicz, ``Deformations de l'algebre de Lie de Poisson d'une variete
symplectique'', Pub. del V congreso de la Ag. de Matematicos de Expresion
Latina (1977) 194-206, Madrid
 1978.

%
\bibitem{LZ2} B.H. Lian and G. Zuckerman,   ``New selection rules and physical
states in 2D gravity: conformal gauge," Phys. Lett.  B254 (1991) 417-.

\bibitem{LZ3} B.H. Lian and G. Zuckerman,  ``Semi-infinite homology and 2D
gravity. I,"  Comm. Math. Phys.  145  (1992), 561-593.

\bibitem{LZ14}
B.H. Lian and G.J. Zuckerman, ``Commutative quantum operator algebras'',
to appear in Journ. Pure Appl. Alg. vol. 100 (1995), q-alg/9501014.

\bibitem{LZ10}
B.H. Lian and G.J. Zuckerman, ``Some classical and quantum algebras'',
hep-th/9404010.

\bibitem{LZ9}
B.H. Lian and G.J. Zuckerman,
``New perspectives on the BRST-algebraic structure in string theory'',
Commun. Math. Phys. 154 (1993) 613-646.

\bibitem{LZ4}
B.H. Lian and G.J. Zuckerman,
``2d gravity with c=1 matter'', Phys. Lett. B266 (1991) 21.

\bibitem{LZ16}
B.H. Lian and G.J. Zuckerman,
``Algebraic and Geometric Structures in String Backgrounds'',
to appear in the Strings 95 proceedings, hep-th/9506210.

\bibitem{Mc}
J. McCarthy,
Int. Journ. Mod. Phys. A Suppl (1995). hep-th/9509134.

\bibitem{M}
G. Moore,
``Symmetries of the Bosonic String S-Matrix'', hepth/93100026.

\bibitem{MMS}
S. Mukherji, S. Mukhi and A. Sen, Phys. Lett 266B (1991) 337.

\bibitem{MVa} S. Mukhi and C. Vafa, ``Two dimensional black hole as a
topological coset model of c = 1 string theory," Nucl. Phys. B407 (1993)
667-705.

\bibitem{MS}
G. Moore and N. Seiberg, ``Classical and quantum conformal field theory'',
Comm. Math. Phys. 123 (1989) 177-254.

\bibitem{nijenhuis}
A. Nijenhuis, ``Jacobi-type identities for bilinear differential concomitants
of certain tensor fields'', Indag. Math., t.17 (1955) 390-403.

\bibitem{OV}
H. Ooguri and C. Vafa, ``Two-dimensional Black Hole and Singularities of
CY Manifolds'', hep-th/9511164.

%
\bibitem{SP}
M. Penkava and A. Schwarz,
``On some algebraic structures arising in string theory'', UC Davis preprint
hepth/9212072.

\bibitem{SS} P. Schaller and T. Strobl, ``A brief introduction to Poisson
sigma-models," hep-th/ 9507020.

\bibitem{Sch}
A. Schwarz,
``Geometry of Batalin-Vilkovisky quantization'',
UC Davis preprint December 92.

\bibitem{TY}
G. Tian and S.T. Yau, ``Complete K\"ahler manifolds with zero Ricci curvature
II'', Invent. math. 106, 27-60 (1991).

\bibitem{Va} C. Vafa, ``Topological mirrors and quantum rings," in Essays on
Mirror Manifolds, ed. S.-T. Yau, International Press, Hong Kong (1992) 96-119.

%
\bibitem{Wi3}
E. Witten,
``Ground ring of the two dimensional string theory'',
Nucl. Phys. B373 (1992) 187.

\bibitem{W2}
E. Witten,
``The anti-bracket formalism'',
preprint IASSNS-HEP-90/9.

\bibitem{W1}
E. Witten,
``Noncommutative Geometry and String Field Theory'',
Nucl. Phys. B268 (1986) 253.

\bibitem{W4} E. Witten, ``Mirror manifolds and topological field theory," in
Essays on Mirror Manifolds, ed. S.-T. Yau, International Press, Hong Kong
(1992) 120-156.

\bibitem{WZ}
E. Witten and B. Zwiebach,
``Algebraic structures and differential geometry in 2d string theory'',
Nucl. Phys. B377 (1992) 55.

\bibitem{WuZhu}
Y. Wu and C. Zhu, ``The complete structure of the cohomology ring and
associated symmetries in $d=2$ string theory'', hepth/9209011.

\end{thebibliography}
\end{document}